\documentclass[lettersize,journal]{IEEEtran}
\usepackage{amsmath,amsfonts}
\usepackage{array}
\usepackage[caption=false,font=normalsize,labelfont=sf,textfont=sf]{subfig}
\usepackage{textcomp}
\usepackage{stfloats}
\usepackage{url}
\usepackage{verbatim}
\usepackage{graphicx}
\def\BibTeX{{\rm B\kern-.05em{\sc i\kern-.025em b}\kern-.08em
    T\kern-.1667em\lower.7ex\hbox{E}\kern-.125emX}}
\usepackage{balance}

\usepackage{algpseudocode}
\usepackage{algorithm}
\usepackage{adjustbox}
\usepackage{booktabs}
\usepackage{multirow, arydshln}
\usepackage{comment}
\usepackage{caption}
\usepackage{subcaption}
\usepackage{amssymb}
\usepackage{amsmath}
\usepackage{mathptmx}
\usepackage{xcolor}
\usepackage{tcolorbox}
\usepackage{lipsum}
\usepackage[noblocks]{authblk}
\usepackage[numbers]{natbib}

\begin{document}
\title{Using LLMs to Automate Threat Intelligence Analysis Workflows in Security Operation Centers}

\author{ PeiYu Tseng; ZihDwo Yeh; Xushu Dai; Peng Liu
\IEEEcompsocitemizethanks{\IEEEcompsocthanksitem PeiYu Tseng, ZihDwo Yeh, Xushu Dai and Peng Liu are with Penn State University, State College,
PA, 16801 (email:pmt5342@psu.edu, doyleyeh@gmail.com, xfd5059@psu.edu , pxl20@psu.edu).
}}

%\thanks{Manuscript created October, 2020; This work was developed by the IEEE Publication Technology Department. This work is distributed under the \LaTeX \ Project Public License (LPPL) ( http://www.latex-project.org/ ) version 1.3. A copy of the LPPL, version 1.3, is included in the base \LaTeX \ documentation of all distributions of \LaTeX \ released 2003/12/01 or later. The opinions expressed here are entirely that of the author. No warranty is expressed or implied. User assumes all risk.}}

\markboth{Journal of \LaTeX\ Class Files,~Vol.~18, No.~9, September~2020}%
{How to Use the IEEEtran \LaTeX \ Templates}

\maketitle

\begin{abstract}
SIEM systems 
are prevalent and play a critical role in 
a variety of analyst workflows in Security Operation Centers. 
However, modern SIEMs face a big challenge: they still cannot 
relieve analysts from the repetitive tasks involved in analyzing  
CTI (Cyber Threat Intelligence) reports written in natural languages. 
This project aims to develop an AI agent 
to replace the labor intensive repetitive tasks involved in analyzing  
CTI reports. The agent exploits the revolutionary capabilities of LLMs (e.g., GPT-4), 
but it does not require any human intervention. 

\end{abstract}


\section{Introduction}
\label{introduction}
Cybercrime has caused significant losses for governments and industries. In 2023, US consumers and businesses lost over \$12.5 billion \cite{IC3}. Among them, 45\% of businesses have experienced a cloud-based data breach or failed an audit \cite{cloud_breach}. This has led companies to place more importance on the preventative attack capabilities of the Security Operations Center (SOC). As the core of a company's security strategy, SOC is entrusted with monitoring and analyzing the organization's network, devices, appliances, and information repositories. Its primary objective is the continual enhancement of the organization's security posture to ensure the protection of its valuable assets. 

During the past two decades, SOC operations have
been continuously evolving, and a clear indicator is that SIEM (Security 
Information and Event Management) systems 
have become prevalent and have evolved significantly. 
SIEM system is equipped with a real-time correlation engine that helps to detect any attacks on company's infrastructure promptly. This engine is responsible for comparing logs with correlation rules stored in the database to check for any matches. When Security analysts identify that the company might be targeted by hacker groups, quickly establishing and updating correlation rules to combat these threats becomes a top priority for the SOC. This proactive step is crucial as the first line of defense, ensuring that the SOC can swiftly respond and mitigate potential security breaches. To establish correlation rules, security analysts typically refer to Cyber Threat Intelligence (CTI). CTI may come from reports by well-known cybersecurity companies such as FireEye and CrowdStrike. It might also come from free security platforms like Mitre ATT\&CK\cite{mitre} or exchanges among security analysts on platforms such as Telegram or X (Twitter).

Today, SIEM systems play a critical role in 
a variety of analyst workflows (e.g., the workflows identifying 
correlated threat patterns, the security monitoring workflows, 
the workflows responding to incidents). And 
a very important SIEM system evolution comes from the rapidly 
emerging threat intelligence market. (The market is 
expected to grow rapidly in the next few years, reaching \$21.92 billion in 
2028 \cite{ctimarket}.)  
This evolution is enabling security operations to shift from 
simple alert systems to advanced mechanisms capable of utilizing 
threat intelligence for purposes such as predictive threat analysis. 

Despite these advancements, modern SIEMs face a big challenge: they still {\em cannot 
relieve} analysts from the labor-intensive {\em repetitive} tasks involved in analyzing  
CTI reports written in natural languages. CTI is often published in reports or blog posts, which requires security analysts to spend a lot of time reading and analyzing. This process also increases the response time to attacks\cite{Benchmark}
Due to this inability, SIEMs have been struggling to   
scale for a large corpus of CTI reports. 
Although existing works (e.g., \cite{BVD15,Gao,Rastogi,Husari2017}) apply machine learning techniques to  
automatically extract ``nuggets'' from security-related documents (e.g., 
policies), the domain-specific AI models are unfortunately shown to be 
inadequate in terms of {\em generalization}.  

In this paper, We seek to develop an AI agent 
to replace the labor-intensive {\em repetitive} tasks involved in analyzing  
CTI reports. The agent exploits the revolutionary capabilities of LLMs (e.g., GPT-4), but it does not require any human intervention. 
With our agent, SOC analysts are relieved from 
repetitive tasks and can spend most of their time on {\em creative} tasks. 

In summary, we we have made the following contributions:
\begin{itemize}
    \item We propose a new AI agent to automate the extraction of important information from CTI reports and generate Regex.
    \item To ensure the accuracy of the generated Regex, we employ a four-step process to filter out potential false positives and false negatives.
    \item Our AI agent can also provide a relationship graph to depict the connections between different CTI within a CTI report.
    \item This project is taking, to our best knowledge, 
    the {\em first} step towards an AI agent that 
    replaces the repetitive tasks without any human intervention. 
    It is also the first work exploiting the revolutionary 
    capabilities of LLMs for making CTI analysis workflows 
    substantially more automated. 
\end{itemize}

\section{Methodology}

\begin{figure*}[!htb]
\centering
\includegraphics[width=\textwidth]{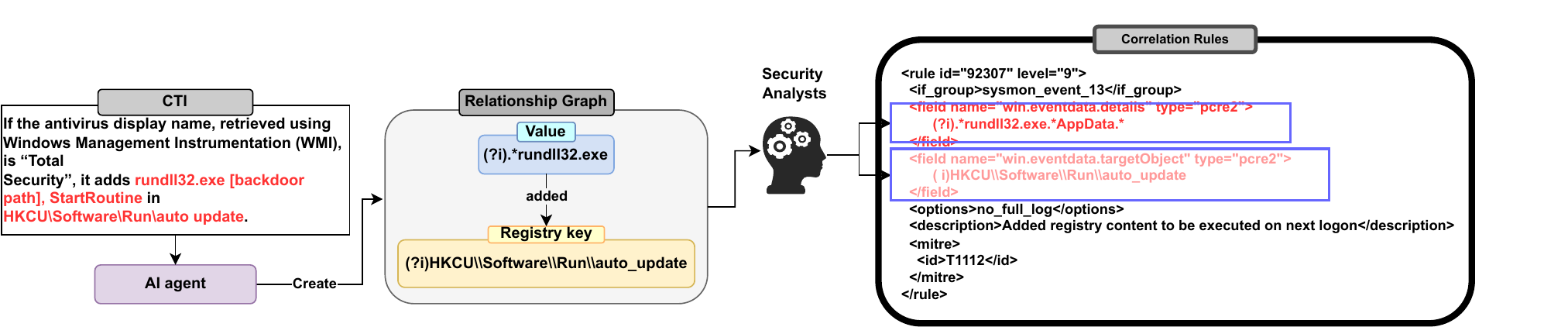}
\caption{Motivating example}
\label{fig:fig4}

\end{figure*}

{\bf Motivating Example.} Figure \ref{fig:fig4} illustrates not only  
a repetitive task conducted by security analysts but also how the task 
is automated by our agent. First, the repetitive task 
is as follows. Given the following paragraph 
from a publicly accessible CTI report \cite{CTI_example}, 
analysts create the two red color fields in 
the SIEM event correlation rule shown on the right side of the figure.  
\begin{small}
\begin{tcolorbox}[colback=gray!5!white,colframe=black!75!black,
         left = 1.0mm, right = 1.0mm, top = 0.6mm, bottom = 0.6mm,title=CTI]
If the antivirus display name, retrieved using Windows Management Instrumentation (WMI), is “Total Security”, it adds \textbf{rundll32.exe [backdoor\_path], StartRoutine} in \textbf{HKCU\textbackslash Software\textbackslash Run\textbackslash auto\_update}.
\end{tcolorbox}
\end{small}
The SIEM rule specifies the following threat pattern: When the Windows machine is infected, a specific malicious registry key 
(i.e., ``HKCU\textbackslash Software\textbackslash Run\textbackslash auto\_update'') is added, and as a result, a 
specific malicious program (i.e., ``rundll32.exe'') will be executed on next logon. This is a repetitive task because none of the two fields belong 
to a new type or name, though the field values cannot be predicted.  
Second, working with our AI agent, analysts {\bf no longer} need to 
digest the CTI report themselves to fulfill this repetitive task. 
Instead, they can directly obtain 
the two field values and their relation from the proposed Relationship Graph, which is shown at the center of the figure. Note that the relationship 
graph is the output of our AI agent. 

%\begin{small}
%\begin{tcolorbox}[colback=gray!5!white,colframe=black!75!black,title=Correlation Rules]
%\begin{verbatim}
%<rule id="92307" level="9">
%  <if_group>sysmon_event_13</if_group>
%\end{verbatim}
%\begin{tcolorbox}[colback=red!10!white,colframe=red!75!black,boxrule=1mm,arc=0mm,outer %arc=0mm,boxsep=0mm,left=1mm,right=1mm,top=1mm,bottom=1mm]
%\begin{verbatim}
%  <field name="win.eventdata.details" type="pcre2">
%    (?i).*\\\\rundll32.exe
%  </field>
%  <field name="win.eventdata.targetObject" type="pcre2">
%    (?i)HKCU\\\\Software\\\\Run\\\\auto_update
%  </field>
%\end{verbatim}
%\end{tcolorbox}
%\begin{verbatim}
%  <options>no_full_log</options>
%  <description>
%    Added registry content to be executed on next logon
%  </description>
%  <mitre>
%    <id>T1112</id>
%  </mitre>
%</rule>
%\end{verbatim}
%\end{tcolorbox}
%\end{small}

{\bf Technical Challenges. } Despite the revolutionary 
capabilities of LLMs, all existing LLMs suffer from 
{\em factual errors}. Even worse, such factual errors 
are {\em unacceptable} to an AI agent that analysts can count on. 
The naive idea of involving human intervention 
can address the factual errors, but will 
unavoidably trap analysts into another repetitive task. 
Without requiring any human intervention, 
the agent must address the following challenges. 
(C1) Although a straightforward prompt can ask 
LLMs to spot the IOCs (Indicators Of Compromise) 
contained in a CTI report, the LLM responses 
must be purified (by the agent) to achieve an acceptable factual error rate.   
(C2) Since SIEM rules must enable searching for more advanced  
(binary) patterns, regular expressions (called RegEx) are employed 
to let SIEM systems have powerful capabilities (e.g., 
partial and case-insensitive matching, binary pattern detection)  
for searching complex data and custom log formats. 
However, 
%since CTI reports are generically applicable to all sorts of enterprise systems,  
the IOCs in CTI reports are written in natural languages and not 
expressed as a RegEx. 
% do {\bf not} hold RegEx. 
Hence, the agent must generate the RegEx.    
To achieve this goal, the agent must correctly distinguish capture groups 
from non-capture groups in IOC strings.
%generic entity names such as \%APPDATA\% must  be replaced with a name {\em specific to} the platforms the analysts guard. LLMs have inadequate ability to do this domain-knowledge-dependent job. 
(C3) 
%SIEM rules are required to hold regular expressions. Although 
After capture and non-capture groups are differentiated, 
LLMs can do a good job of generating the RegEx. However, a mechanism must be 
developed to automatically test the LLM outputs. 
% Otherwise, the agent may again fail to achieve an acceptable factual error rate.
(C4) Although LLMs can help identify the dependencies between 
the extracted IOCs, the factual errors in LLM responses may 
result in unacceptable Relationship Graphs. 
It is challenging to automatically ``filter out'' the factual errors.

%--------------------
%\vspace*{1mm}
\subsection{Overview of the AI Agent}

\begin{figure*}[!htb]
   
    \centering
    \includegraphics[width=\textwidth]{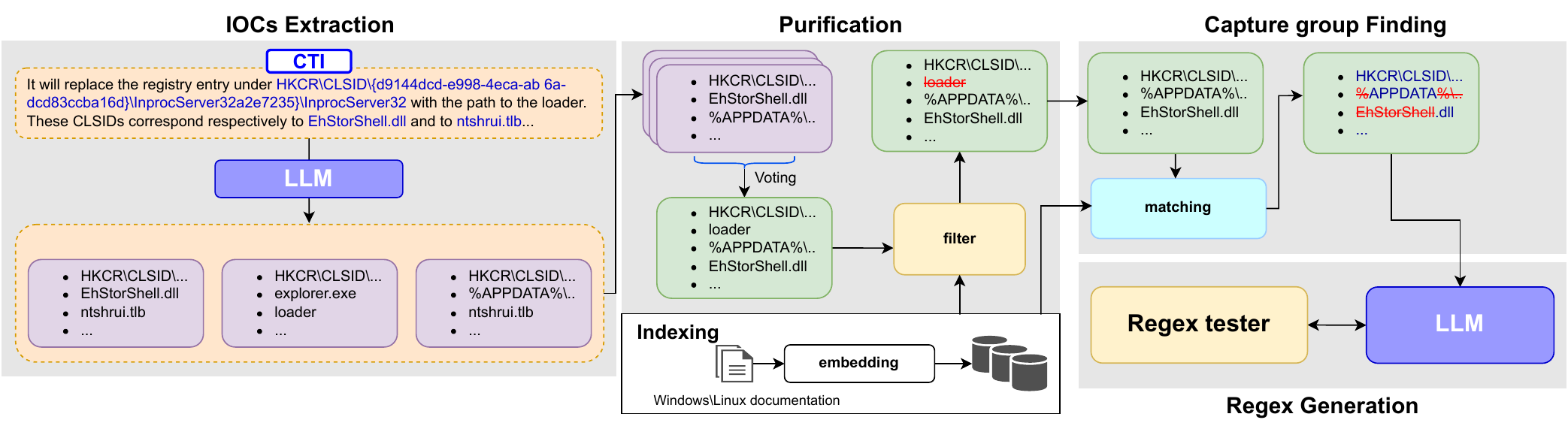}
    \caption{Workflow of the proposed AI agent: the first half}
    \label{fig:fig2}
    
    \end{figure*}

\begin{figure*}[!htb]
    \centering
    \includegraphics[width=\textwidth]{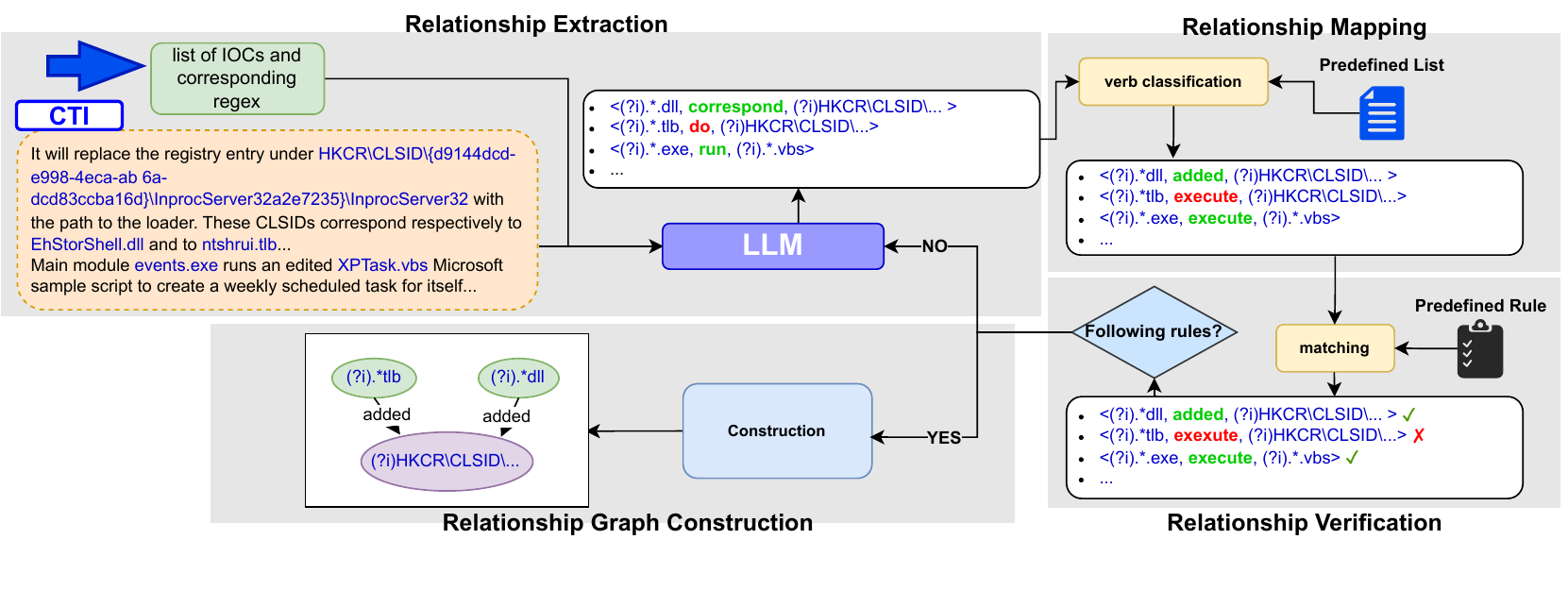}
    %\vspace*{-3mm} 
    \caption{Workflow of Relationship Graph Construction}
    \label{fig:fig1}
    \end{figure*}

Figure \ref{fig:fig2} shows the first half of our agent's workflow, while Figure \ref{fig:fig1} shows the second half. 
%Let's illustrate the workflow through the above motivating example. 
In Step 1, instead of feeding the entire CTI report 
into the LLM (i.e., GPT-4), our agent asks the LLM  
to concentrate on each individual paragraph and extract IOCs. 
In Step 2, in order to address challenge C1, the LLM responses are purified 
through a combination of voting (among several LLM runs) and 
retrieval-augmented filtering. 
In Step 3, in order to address challenge C2, a retrieval-augmented matching 
mechanism is proposed to 
distinguish capture groups from non-capture groups in IOC strings. 
In Step 4, our agent asks the LLM to generate the RegEx 
used in SIEM rules and addresses Challenge 3 through a RegEx tester. 
In Step 5 (see Figure \ref{fig:fig1}), our agent 
leverages the LLM to identify the   
dependencies between the extracted IOCs.  
In Step 6 and Step 7, in order to address Challenge C4, 
our agent categorizes 
%(i.e., identify  the type of each dependency) 
and verifies each identified dependency relationship, 
respectively. In Step 8, our agent constructs the proposed
Relationship Graph (see Figure \ref{fig:fig4}).

%For example, given the following paragraph from CTI \cite{CTI_example}, the AI agent will first identify two IOCs 'rundll32.exe [backdoor path], StartRoutine' and 'HKCU\textbackslash Software\textbackslash Run\textbackslash auto\_update'. The former will be labeled as a Value, while the latter will be tagged as a Registry key. The agent will then generate corresponding regexes, '(?i).*rundll32.exe' and '(?i)HKCU\textbackslash \textbackslash Software\textbackslash \textbackslash Run\textbackslash \textbackslash auto\_update'. Since the Value 'rundll32.exe' is added to the Registry key 'HKCU\textbackslash Software\textbackslash Run\textbackslash auto\_update', the AI agent will link these two IOCs and mark the connection as 'added'. 
%As show in figure \ref{fig:fig4}, our AI agent will identify which IOCs are present in this text paragraph and what they belong to, such as a file or registry key, then generate regex for them. The AI agent will then further identify the connections between these IOCs and create a relationship graph. This AI agent aim to help security analysts understand how to effectively use the information in CTI to establish reliable rules and reduce false positives in threat detection.

\subsection{Design Details of the AI Agent}

%In this section, we will first demonstrate how we use LLM to extract Indicators of Compromise (IOCs) from CTI and generate the regular expression for the correlation rule. Then, we will use our framework to create a relationship graph among the IOCs. 

\textbf{IOCs Extraction}: 
Researchers have studied how to extract information ``nuggets'' from CTI reports  \cite{Gao,Rastogi,Husari2017}. 
However, existing works rely on pre-trained Named Entity 
Recognition (NER) models, which are very limited in identifying new 
named entities in evolving attack techniques. 
Recently, the powerful capabilities of LLM in 
%extracting information have been demonstrated, particularly in 
the domain of recognizing specialized terminology have been demonstrated 
(e.g., \cite{xu2023large}). 
In this project, instead of using domain-specific NER models, 
we will utilize LLMs to identify IOCs. 
In particular, our agent divides each CTI report into a set of paragraphs, each 
containing 3-4 sentences. For each paragraph, we will utilize LLMs to  
%words are nouns, and will further instruct the LLM to 
identify the nouns that are relevant to IOCs, such as 
filenames, command lines, and registry keys. 

%\vspace*{1mm}
\noindent \textbf{Purification}: 
%Current LLMs may not fully understand human instructions, which can lead to false positives. 
In the previous step, when our agent asks an LLM to identify the 
nouns corresponding to filenames or command lines, there is a certain probability that the LLM provides completely irrelevant answers. 
To address this issue, our agent will adopt a ``majority voting'' mechanism to filter out 
potential factual errors. 
% false positives. 
Our agent will have the LLM analyze the same paragraph multiple times and 
conduct majority voting against the responses. 
%observe which answers appear more frequently than a predetermined threshold. 
Only the responses that meet or exceed a pre-determined voting threshold 
will be retained. 
%while those with lower frequencies will be considered false positives and discarded. 
Since the retained responses may still suffer from factual errors, 
% To further ensure the accuracy of the LLM results, 
our agent will leverage the 
concept of Retrieval-Augmented Generation (RAG) \cite{gao2024retrievalaugmented} to 
filter and adjust the retained responses. 
First, we gather domain knowledge (e.g., 
PowerShell commands, default file folder names) 
about the enterprise system guarded by SOC from 
Windows or Linux operating system documentation \cite{WindowsOS,Linux}. 
The knowledge will be encoded into vectors and stored in a vector database. 
Next, our agent will match the retained LLM responses 
% format of the IOCs to see if they align with the information in 
with the vector database. 
For example, check whether a filename identified by the LLM 
exists as a path (or includes an extension) stored in the database; 
check whether a command line identified by the LLM 
follows the command formats used by Windows or Linux.

%\vspace*{1mm}
\noindent
\textbf{Capture Group Finding}: %Since CTI is written for human reading, the IOCs mentioned therein sometimes cannot be directly used to generate regex. For instance, some filenames are written as environment variables in CTI, like "$\%APPDATA\%$", and in such cases, they need to be converted back to complete system paths. Additionally, 
When creating RegEx for a purified IOC string, our agent must  
distinguish capture groups from non-capture groups inside the string. 
Typically, capture groups include parts of the string that attackers cannot easily 
modify, such as arguments in command lines, system program names, or 
default file directories. 
On the other hand, non-capture groups are parts of the string that 
attackers can change, such as file or variable names in command lines.
For example, ransomware uses the following command line \cite{CTI_example3} to delete volume shadow copies: 
        \begin{small} 
        \begin{tcolorbox}[colback=gray!5!white,colframe=black!75!black,
                left = 1.0mm, right = 1.0mm, top = 0.6mm, bottom = 0.6mm,title=Command line]
            cmd.exe /c \%System\%\textbackslash wbem\textbackslash WMIC.exe shadowcopy where ``ID='GUID''' delete
        \end{tcolorbox}
        \end{small}
Here, ``cmd.exe'', ``WMIC'', ``shadowcopy'', 
and ``delete'' correspond to capture groups, 
%are part of the capture group, 
while ``GUID'' belongs to a non-capture group. 
%To identify which substrings are capture groups and non-captured groups, we will split the string into multiple substrings based on its format. 
Accordingly,  
our agent will first split the string into several substrings. 
For command lines, spaces are used as boundaries to extract substrings; 
for paths, slashes will be utilized as delimiters. 
Second, our agent converts the substrings into embedding vectors, 
after which a comparison will be conducted with the vector database. 
Matched substrings are considered as capture groups. 
%conversely, if no matching strings are found, they 
The remaining substrings are categorized as non-capture groups. 
%\noindent \textbf{Regex Generation}: 
In the prompt, our agent will annotate which parts of the IOCs are capture groups 
and which parts are non-capture groups. 
Since the LLM responses could suffer from invalid
RegEx format, our agent will use a RegEx tester to perform 
step-by-step testing and provide feedback to the LLM, allowing it to 
further refine its responses until the RegEx is completely valid. 

%\vspace*{1mm}
\noindent \textbf{Relationship Extraction}: 
To construct the proposed relationship graph, our agent must identify 
%which entities, specifically the IOCs, have 
the dependencies between the IOCs extracted in the previous steps. 
For this purpose, our agent will first ask the LLM to 
revisit the original paragraph (see Step 1) 
and identify all the nouns and verbs. 
%and determine whether there is a dependency relationship between pairs of nouns. 
For example, let's examine the following paragraph from CTI report \cite{CTI_example2}.  
        \begin{small}
        \begin{tcolorbox}[colback=gray!5!white,colframe=black!75!black,
                left = 1.0mm, right = 1.0mm, top = 0.6mm, bottom = 0.6mm,title=CTI]
            \textbf{NSC Press conference.exe}, acts as a \textbf{dropper}. Whether the \textbf{dropper} found a document to open or not, it will proceed to the next stage – drop the backdoor to \textbf{C:\textbackslash users\textbackslash public\textbackslash spools.exe} and execute it.
        \end{tcolorbox}
        \end{small}
Our agent will first ask the LLM to identify all nouns and their paired nouns, 
such as ``NSC Press conference.exe - dropper'',  ``dropper - document'', and 
``dropper - C:\textbackslash users\textbackslash public\textbackslash spools.exe''. 
Then, our agent will compare each pair of nouns with the   
IOCs extracted in the previous steps. 
%this with the results from IOC extractions. 
Since none of the nouns in ``dropper - document'' is an IOC, 
this pair will be discarded. 
Afterward, our agent uses the extracted verbs to determine whether a noun refers to another. 
For instance, since the relationship between ``NSC Press conference.exe'' and ``dropper'' 
is ``act'', ``dropper'' is a pronoun referring to ``NSC Press conference.exe''. 
In this way, all pronouns in the retained pairs will be replaced with the corresponding IOCs. 
Finally, each retained pair corresponds to one identified dependency 
and it is not difficult to find the associated verb. 

%\vspace*{1mm}
\noindent
\textbf{Relationship Mapping}: CTI reports 
use a variety of verbs to describe the relationships between IOCs, 
and there is no standard format. For instance, when describing the action of malware creating a file, sometimes ``drop'' is used, while other times ``establish'' is employed. 
To avoid confusing (inexperienced) analysts, 
% facilitate the understanding of relationships between IOCs for security analysts, it's essential to 
our agent will standardize these verbs. 
In particular, our agent will use a mapping table 
we have previously created to categorize each verb. 
%the extracted verbs to corresponding relationship based on a list, this list will include several IOCs relationship verbs representing data flows, such as "create", "write", "load", "read", and others. 
For example, ``drop'' or ``establish'' would be mapped to the 
``create'' category; 
``change'' and ``edit'' would be mapped to the ``write'' category. 
%\vspace*{1mm}
\textbf{Relationship Verification}: 
%To ensure that the relationship graph accurately describes the relationships between the IOCs, we will verify the relationship verbs between IOCs to see if they are reasonable. 
To further reduce factual errors, our agent will 
verify the relationship verbs between IOCs. 
For example, IOCs categorized as ``Registry key'' should never have a ``create'' 
relationship pointing to a ``file'' or ``process''. 
Once an incorrect relationship is discovered, our agent will locate the 
paragraph where these two IOCs appear and 
have the LLM re-identify their relationship.
%\vspace*{1mm}
\textbf{Relationship Graph Construction}:
%We will construct relationship graphs by utilizing all available IOCs in conjunction with their relevant relationships. In the graph, the entities represent the IOCs and their corresponding regexes, while the edges represent the relationships between the IOCs and the direction of the information flow.
After all the relationships are identified, the relationship graph is 
straightforward: each node is an IOC expressed as a RegEx;  
and each edge is a dependency relationship identified in the previous step. 
Directed edges capture the direction of the information flow between IOCs. 

\section{Evaluation}

We have conducted experiments on 50+ CTI reports. In these experiments, the LLM identified over 2,900 potential IOCs. Through purification, we found around 2,300 valid IOCs, 
including filenames, domain names, hash values, IP addresses, command lines, registry keys, and values.  
Among these, IOCs such as hash values, IP addresses, and domain names 
% at the bottom of the Pyramid of Pain 
make up 70\% of the total. 
Moreover, our agent generated about 2,200 RegEx. 
The reason for having fewer RegEx than valid IOCs is due 
to the appearance of similar IOCs across multiple CTI reports, such 
as the ``AppData'' folder or registry keys like ``HKLM\textbackslash Software\textbackslash Microsoft\textbackslash Windows\textbackslash CurrentVersion\textbackslash Run''. 
These IOCs have the same capture group.
%therefore the generated regular expressions will also be the same. 
%Moreover, 
Compared to the manually identified ground truth, our AI agent only failed to identify 3\% of the IOCs.  
%belonging to network or host artifacts. 

%\input{tex/7-related}
\section{Conclusion}
This paper presents a pioneering effort to develop an AI agent designed to automate the labor-intensive and repetitive tasks associated with analyzing CTI reports. By leveraging the advanced capabilities of LLMs, our AI agent can accurately extract important information from large volumes of text and generate Regex to help SOC analysts accelerate the process of establishing correlation rules. In addition, the AI agent can extract relationships between CTI and automatically create relationship graphs, helping SOC analysts understand the attack models of adversaries. With the aid of automated Regex generation and relationship graph creation, SOC will be able to reduce response times to attacks.

\bibliographystyle{abbrv}
\bibliography{refs}

%Biography text here without a photo.

%\begin{IEEEbiography}[{\includegraphics[width=1in,height=1.25in,clip,keepaspectratio]{images/state.pdf}}]{IEEE Publications Technology Team}
%In this paragraph you can place your educational, professional background and research and other interests.\end{IEEEbiography}

\end{document}